Almila Akdag Salah, Cheng Gao, Krzysztof Suchecki, e-Humanities Group, KNAW, Amsterdam, Netherlands
Andrea Scharnhorst, DANS, KNAW, Amsterdam, Netherlands
Richard P. Smiraglia, University of Wisconsin, Milwaukee

# The evolution of classification systems: Ontogeny of the UDC

**Abstract**
To classify is to put things in meaningful groups, but the criteria for doing so can be problematic. Study of evolution of classification includes ontogenetic analysis of change in classification over time. We present an empirical analysis of the UDC over the entire period of its development. We demonstrate stability in main classes, with major change driven by 20th century scientific developments. But we also demonstrate a vast increase in the complexity of auxiliaries. This study illustrates an alternative to Tennis' "scheme-versioning" method.

## 1. Introduction

Beghtol says that "to classify means to put things into meaningful groups" (2010, 1045). Defining the "things" that are to be classified, determining their attributes and appellations, and drawing inclusion-exclusion boundaries are all dependent on the social construction of knowledge. A classification, being more than a simple tree or simple set of mutually exclusive boxes, is rather more like an evolutionary tree, in that the relative location of species on the tree is affected not only by empirical knowledge of the species but also by the socio-historic character of knowledge in general as it grows unevenly over time. Tennis has written about this evolution in two ways, as subject ontogeny (2002) and as scheme versioning (2007). In his paper on subject ontogeny, Tennis refers to the evolution of classification as the revision of a knowledge organization system (KOS) as "a revised scheme absent of any record of what the knowledge landscape looked like before" (2002, 54). His analogy adds the concept of space-time, as the KOS which undergoes constant revision becomes perceived as static against a constantly-shifting landscape, which itself appears to have remained static. This process is the phenomenon of instantiation (Smiraglia 2005, 2008), in which information objects are realized at various points along a temporal trajectory. Tennis calls this junction between instantiations of KOS and spacetime "the dimensionality of classification" (Tennis 2002, 56). In each case, the failure to recognize earlier or later instantiations, as well as earlier or later knowledge landscapes against which the instantiation is realized, leads to chaotic interpretation. It is this chaotic interpretive line that Tennis refers to as "scheme versioning," as he offers an empirical approach to locating individual species against these shifting space-time backdrops. The purpose of this paper is to advance understanding of scheme versioning by demonstrating the evolution of a constantly instantiating KOS, namely the Universal Decimal Classification (UDC) system. Elsewhere we have used our evolutionary analysis of the UDC as a stable reference system over against the volatility of the knowledge landscape itself represented by the constantly shifting knowledge network in Wikipedia (Akdag Salah et al. 2012; Scharnhorst et al. 2011). In this paper we use the UDC as a case study to demonstrate how we have analyzed the instantiating evolutionary tree of the UDC over time.

## 2. Data and method

The data that we analyze in this paper stems from different years and formats. We have obtained the first edition of UDC as a printout in French (International Federation for Documentation, 1905). This edition contained only about 400 records, a rough draft of the main classes, and the second classes only. We manually entered the information from the printed edition into an EXCEL file. The rest of our data comes from the UDC archives (the courtesy of UDC Consortium, The Hague). Currently, we have the master reference files of UDC from the years 1994 (61563 records), 1997 (61000 records), 1998 (60685 records), 2005 (66495 records), 2008 (68193 records) and 2009 (68551 records).

In order to analyze the data we decided to apply various visualizations, and chose MagnaView as our visualization software. For this purpose, we have written a Java-based program to convert the UDC Master Reference text file into a tab delimited text file that can be read by MagnaView. In the MRF each record contains one unique UDC number (Strachan and Oomes 1995). "The UDC Master Reference File (MRF) is a database that contains the UDC schedules together with records needed for administration, maintenance and archiving".[1] The process of conversion needed some decisions of how to handle the UDC data. First we parsed the UDC number and cut the string into digits. Hereby, we followed a relative simple method of splitting and counting UDC numbers. The approach is based on the fact that UDC numbers are constructed in a way that directly reflects the class structure, including the fact that the main class number comes first. To investigate the auxiliary usage in UDC we added two extra fields in the MagnaView datafile structure and assigned each record its auxiliary type and extracted the auxiliary part.

## 3. Results
### 3.1 Changes in the main classes

The UDC has ten main classes (Harper 1954) at the top level of its tree structure. In Figure 1 (a) we illustrate the distribution of these ten classes at two different points in time, specifically and respectively 1905 and 1994. Over the century of development two main changes took place: 1) the second level category "[01] Bibliographie" became a part of the main class scheme, and was expanded to include not only library studies, but 'Science and Information Organization' in general. A new addition to this class is 'Computer Science'; and 2) the class "[4] Philology. Linguistics" was vacated. This was accomplished by moving 'Linguistics' to "[8] Literature", and removing the term 'Philology' from the main classes. To capture more fundamental changes between the main classes during the century, we have decided to include ''[01] Bibliographie'' as a main class, and highlighted it with the same color of [0] Science and Information Organization. There are two simple reasons for our decision: first, we want to show the expansion of [01] to [0] in numbers, and second, [01] on itself was not a subclass of any other main class in 1905. Otherwise the main class structure of the UDC remained stable, and saw only changes at lower levels. These changes, though not interfering with the main class structure, can be quite severe themselves. The annual issues of the Extensions and Corrections to the UDC list changes such as: entry and exit of UDC numbers, changes in the description of UDC numbers, shifting entries among classes and so on.

Since the launch of the 1993 Master Reference File the revision process is under supervision of the editorial board and its editor in chief. Revisions on specified topics are commissioned and updated tables are submitted to specialists for quality control prior publication. With our visual analysis we highlight some of the traces of those changes echoing the complex editorial process around the UDC in a period of relative stability.

In relation to 1905, all main classes show an enormous growth, however, if we compare the changes not in pure numbers, but in percentages, we see that [0] has shrunk considerably more than other classes. Actually, accept [5] and [6], all the classes' percentages decreased.

---

[1] See http://www.udcc.org/mrf.htm for more details.

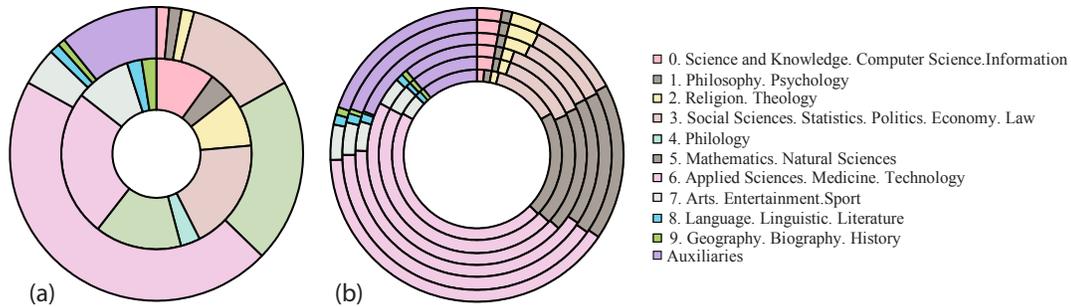

**Figure 1: (a) Distribution of main UDC classes, inner ring 1905, outer ring 1994. (b): Distribution of main UDC classes in 1994 (most inner ring), 1997, 1998, 2005, 2008 and 2009 (most outer ring).**

The first editions of UDC attempted to encompass universal knowledge, and that was reflected in the (comparatively) even distribution of the main classes. A comparison of the composition of the UDC between 1994 and 2009 reveals a relative stability in the share of UDC numbers across disciplines (see Figure 1 (b)). The modern UDC, as it is obvious in the growth of the proportions, is mainly occupied with natural [5] and applied sciences [6]. This might reflect the increasing importance of science and technology in society, but it might also be a consequence of the development of libraries, and the bias in library collections, for which UDC is mainly used today. What also caught the eyes is the apparent (relative) increase of use of auxiliaries between 1998 and 2005. Obviously, auxiliaries play an important role in the adaptation process of the UDC.

### 3.2 Changes in Special Auxiliaries

UDC contains two types of auxiliaries, common and special auxiliaries. Special auxiliaries provide a means to express complexity in proscribed specific classes (thus each is "special" to the class or classes for which it was been designated). As we have mentioned before, the usage of special auxiliaries differ among the main classes. UDC contains different type of special auxiliaries used with different connotations.

For the sake of simplicity, we have disregarded these differences, and treated all special auxiliaries as the same. We are more interested in which main classes special auxiliaries are applied and to what extent. Figure 2 (a) shows the distribution of them among the main classes, and their changes over the years. As expected, main class [6] has the most special auxiliaries, even though their number is on the decrease from 9613 in 1994 to 9442 in 2009. Here the rate of decrease follows a soft curve, in contrast to a few jumps we observe in other main classes. For example, class [2], i.e. Religion, has an unprecedented change in the special auxiliary numbers from 1998 to 2008: from 239 records to 2168 records. If we consider the total record amount class [2] covers, we see that this class has seen a substantial change over time: In 1998, [2] contained only 935 records, which has arisen to 2419 records in 2008. The transformation of special auxiliaries here is paramount; in 1998 only 14.87% of the class [2] was occupied by special auxiliaries, whereas in 2008 this number has risen up to 89.6 %.

Figure 2: (a) Distribution of special auxiliaries among the main classes over the years.
(b): Percentage of special auxiliaries to main classes, and their changes over time

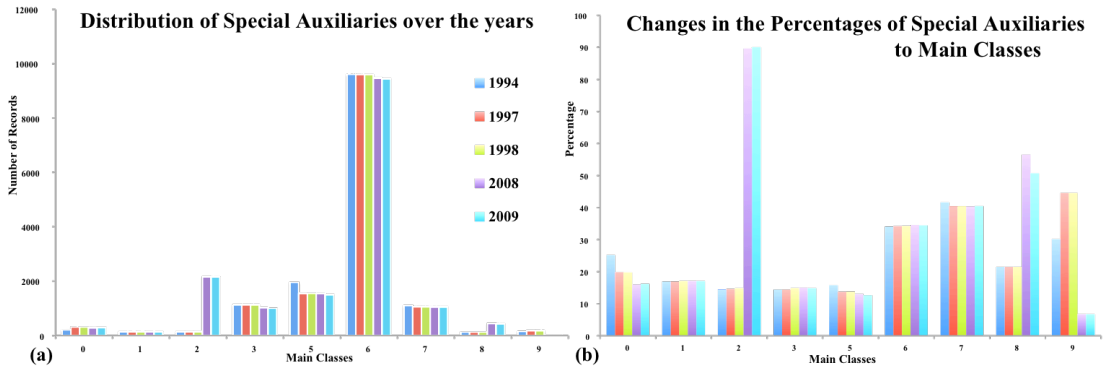

Similarly, we observe sharp fluctuations among the classes [5], [8] and [9]. To understand the importance of these, it is best to compare the changes in connection with the changes in the main classes themselves. Figure 2 (b) shows the changes of special auxiliaries, but this time, the y-axis of the chart reflects the percentage of the special auxiliary record numbers to the total number of any given main class. Thus we see that after 2008, the main class [2] is been transformed almost entirely; now % 90.07 of it is only special auxiliaries. The main classes [8] and [7] also have a considerable amount of special auxiliaries (50.7% and 40.48 % respectively), whereas in the main class [9] the use of special auxiliaries has been reduced from 44. 44.58 % in 1998 to 6.75 in 2009.

The impression we get from the Figure 2 (a) is that the most special auxiliaries appear in the main class [6]. When we speak of the number of records a given class covers, this impression is correct. However, if we want to see the percentage of each class occupied by special auxiliaries, we must consult the Figure 2 (b).

### 3.3 Changes in Common Auxiliaries

Common auxiliaries are used to express language, form or cultural origin; they are called "common" because they can be used with any number where appropriate (thus they are common to the entire UDC). Common auxiliaries in the UDC are on the increase. As also indicated in Figure 1 (b), in 1998 the number of common auxiliary records was 6812, whereas today, this number has reached a total of 13562. So, the use of auxiliaries grows over-proportional in the UDC evolution, especially the common auxiliaries. In comparison to this sharp increase, all other fluctuations among the main classes seem to be trivial.

In order to understand why common auxiliaries occupy so much more space in UDC, and how they change the structure of the UDC tree itself, we need to further analyze this phenomenon. Figure 3 (a) displays the distribution of common auxiliaries over time. Since our data stems from the Main Reference File, and does not reflect the actual usage of UDC, common auxiliaries such as 'a' and 'b' have only one record, and that is the description of how these auxiliaries should be used. These auxiliaries are above all for combining different UDC main records, i.e. records from the Main Table, and thus in our analysis we cannot trace their effects. A similar common auxiliary is the type 'h', which incorporates non-UDC sources into UDC records. This auxiliary also come to life only in actual usage, and does have only one record in all the MRFs.

The rest of the common auxiliaries need to be assessed further: Among these, 'e', the auxiliary of the place, is the most used one, as well as the one with the highest increase in numbers. The second most used common auxiliary is the type 'c', auxiliary of the language. However, the second auxiliary that has a sharp increase is the auxiliary of persons and materials, i.e. the auxiliary 'k'. The auxiliary of form (type 'd') remains stable, and the auxiliary of time (type 'g') shows only a small increase. Thus, we can conclude that the main changes in the record numbers of the MRF files from 1993 to 2009 have taken place among auxiliaries, and among them, the reason of this sharp increase comes from the additions to the auxiliary 'e' and 'k'.

**Figure 3: (a) Distribution of common auxiliaries over the years. (b) Changes in the record number of main classes from 1993 to 2009.**

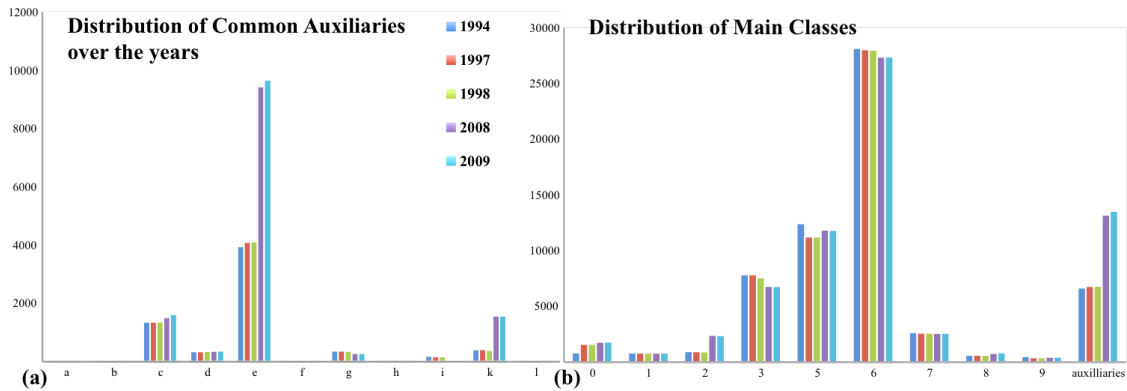

### 4. Conclusions

With this analysis we examine some features of the dynamics of a knowledge organization system in time. Our explorations are of importance along different dimensions: theoretical debates inside of knowledge organization and information sciences; the UDC consortium; and aspirations to better understand and manage complex knowledge spaces across many scientific disciplines.

### 4.1 Ontogenetic analysis

Tennis (2007) proposed a specific schema for ontogenetic analysis of individual concepts as they evolve over time in a given KOS. Our approach has been different, in that we have chosen to analyze the evolving structure of the UDC against the socio-cultural knowledge landscape of the 20th century in which it developed.

For the most part, change in the UDC reflects well the growth of knowledge over the past century. Interestingly, we have observed a vast increase in the auxiliaries. This includes the enumeration of peoples and places in the common auxiliaries, reflecting increased cultural sensitivity in the UDC, but also special auxiliaries which increased dramatically in number in applied sciences and religion. This is a shift in what Tennis called "dimensionality," as the UDC grew not in its extension, but rather in its intension, as the branches of the initial tree were expanded with auxiliaries. Here, Beghtol's inclusion-exclusion boundaries shift in dimension across spacetime. Hence we understand there is not at any point a single UDC, but rather, at every point, there is the UDC in spacetime.

Our results demonstrate the efficacy of this analytical approach to ontogenetic analysis. We believe that this approach, combined with Tennis' suggested approach to scheme versioning provides a useful methodological triangulation that incorporates empirical knowledge about the evolving instantiating, living, UDC.

### 4.2 The UDC Consortium

Not surprisingly with a history of more than 110 years, the UDC has gone through many transformations in itself, but also in the ways it is managed. By 1980s, UDC had many different versions in various languages, and its full version exceeded 200,000 records. To produce editions of the UDC in different languages which can be applied in collections around the globe is one focus point of the editorial board. Another one is of course the adaptation of the UDC to the growing and changing body of knowledge. A great attention and care is devoted to first make this changes and with them keep the UDC as a living knowledge order system (and not a historic one); but second also to do it with care to guarantee a certain continuity so that collections don't have to re-index their holdings all over the time. Not surprisingly, the growth and change in the UDC did not always follow a strict strategy with some problematic results, as the editorial board pointed out in several publications themselves. For example, the original 19th century knowledge-base had overlapping concepts from the 20th century. As one solution a medium version is generated by a selected editorial board (Slavic et. al., 2008). This was formed in 1989 from the 1985 English edition, which was already in machine-readable format (McIlwaine 1997). The resulting updated Master Reference File was launched in 1993.

Our analysis traces some of the changes in a period the editorial regime of the UDC was relative stable. However, the changes we reported can only be properly understood knowing the context and history of editorial changes in the UDC. Sudden changes in records in some classes can be related to specific concerted action of their profoundly renewing. Auxiliaries are used to avoid repetition of subclasses through different main classes[2], but seem also have to be used to add diversity and flexibility without disturbing the overall structure to much. The different editions of the "Guidelines to the UDC" and further publications from the consortium (e.g., Riesthuis 1998) and others contain such context information. Yet a comprehensive history of the UDC including elements of statistical and visual analytics is pending. Nevertheless, we believe that this kind of statistical analysis can also trigger further debates about the UDC as complex KOS. (Scharnhorst et al 2012)

**4.3 Evolving knowledge spaces**

The UDC as a complex KOS is rather a language to describe objects in a controlled way than a hierarchical tree which allows to (co-)allocate objects. The possibilities to combine UDC numbers directly but also their indirect coupling via shared auxiliaries makes the UDC to a complex network. (Barabasi 2002, Scharnhorst 2003) With this paper we only looked at the surface of what is possible in terms of analyzing the evolving network of the UDC. Moreover, we analyzed the UDC only in terms of its Master Reference File appearance. One could say that this is the "designed" side of the UDC – or the UDC master plan. The application of the UDC in libraries and museums creates another rich set of UDC expression in collections. In other words, via a shared KOS as the UDC we can analyze the evolution of knowledge as recorded in space-time specific objects in collections, which are also divers in time and space. Insights in the dynamics of knowledge production and its organization can help us to better understand and navigate through large-scale knowledge spaces as available via the web.

---

[2] Personal communication Gerhardt Riesthuis.